\documentclass[aps,noshowpacs,twocolumn]{revtex4}
\usepackage{amsmath}
\usepackage[bookmarks=false,colorlinks,linkcolor=blue,citecolor=blue,urlcolor=blue]{hyperref}
\usepackage{graphicx}
\usepackage{color}
\usepackage{epstopdf}
\begin{document}
\title{Investigation of $cs\bar c\bar s$ tetraquark in the chiral quark model}

\author{Yifan Yang, and Jialun Ping\footnote{Corresponding author: jlping@njnu.edu.cn}}
\affiliation{Department of Physics, Nanjing Normal University, Nanjing 210097, P.R. China}

\begin{abstract}
Inspired by the recent observation of exotic resonances $X(4140)$, $X(4274)$, $X(4350)$, $X(4500)$ and $X(4700)$
reported by several experiment collaborations, we investigated the four-quark system $cs\bar c\bar s$ with quantum
numbers $J^{PC}=1^{++}$ and $0^{++}$ in the framework of the chiral quark model. Two configurations, diquark-antidiquark
and meson-meson, with all possible color structures are considered. The results show that no molecular state
can be formed, but the resonance may exist if the color structure of meson-meson configuration is $8\otimes8$.
In the present calculation, the $X(4274)$ can be assigned as the $cs\bar c\bar s$ tetraquark states with
$J^{PC}=1^{++}$, but the energy of $X(4140)$ is too low to be regarded as the tetraquark state. $X(4350)$ can be a
good candidate of compact tetraquark state with $J^{PC}=0^{++}$. When the radial
excitation is taken into account, the $X(4700)$ can be explained as the $2S$ radial excited tetraquark state
with $J^{PC}=0^{++}$. As for $X(4500)$, there is no matching state in our calculation.
\end{abstract}

\maketitle

\section{INTRODUCTION}
Recently, several exotic resonances were observed in the invariant mass distribution of $J/\psi\phi$.
In 2009, the CDF Collaboration found the $X(4140)$ with mass $M=4143.0\pm2.9\pm1.2$ MeV and width
$\Gamma=11.7^{+8.4}_{-6.7}\pm3.7$ MeV in $B^+\rightarrow J/\psi\phi K^+$ decay~\cite{1}. In 2010,
a narrow resonance $X(4350)$ with mass $M=4350.6^{+4.6}_{-5.1}\pm0.7$ MeV and width
$\Gamma=13^{+18}_{-9}\pm 4$ MeV was reported by Belle Collaboration in $\gamma\gamma\rightarrow J/\psi\phi$
process and the possible spin parity is $J^{PC}=0^{++}$ or $2^{++}$~\cite{2}. A few years later,
the exotic resonance $X(4140)$ was observed by some other collaborations including LHCb, D0, CMS, and
BarBar~\cite{3,4,5,6}. In 2011, another resonance $X(4274)$ with mass $M=4274.4\pm1.9$ MeV and width
$\Gamma=32.3\pm7.6$ MeV was observed by the CDF Collaboration in $B^+\rightarrow J/\psi\phi K^+$ decay
with $3.1\sigma$ significance~\cite{7}.
In 2016, the LHCb Collaboration performed the first full amplitude analysis of the $B^+\rightarrow J/\psi\phi K^+$
process and the existence of the $X(4140)$ and $X(4274)$ was confirmed. Their quantum numbers are fixed to be
$J^{PC}=1^{++}$~\cite{8}. At the same time, the Collaboration observed another two resonances, $X(4500)$ and $X(4700)$
with $J^{PC}=0^{++}$. Their masses and decay widths have been determined as~\cite{9}
\begin{eqnarray}
M_{X(4500)} & = & (4506\pm11^{+12}_{-15}) ~\mbox{MeV}, \\
\Gamma_{X(4500)} & = & (92\pm21^{+21}_{-20}) ~\mbox{MeV}, \\
M_{X(4700)} & = & (4704\pm10^{+14}_{-24}) ~\mbox{MeV}, \\
\Gamma_{X(4500)} & = & (120\pm31^{+42}_{-33}) ~\mbox{MeV}.
\end{eqnarray}

With the discovery of these exotic resonances, many theoretical work has been performed, such as approaches based on quark models~\cite{10,11,12,13,14}, QCD sum rules~\cite{15}, etc. In the framework of relativized quark model, the $X(4140)$ can be
regarded as the $cs\bar c\bar s$ tetraquark ground state, and the $X(4700)$ can be explained as the $2S$ excited tetraquark
state~\cite{10}. Based on the simple color-magnetic interaction model, possible ground $cs\bar c\bar s$ tetraquark states in
the diquark-antidiquark configuration have been investigated, and the interpretation of $X(4500)$ and $X(4700)$ needs
orbital (radial or angular) excitation~\cite{11}. Deng {\em et al.} investigated the hidden charmed states in the framework of the
color flux-tube model, and they found the energy of the first radial excited state $(cs)(\bar c\bar s)$ with $2^5D_0$ is
in full accord with that of the state $X(4700)$~\cite{12}. In a simple quark model with chromomagnetic interaction,
Stancu suggested that the $X(4140)$ could possibly be the strange partner of $X(3872)$ in a tetraquark
interpretation~\cite{13}. Ortega {\em et al.} claimed that the $X(4140)$ resonance appears as a cusp in the $J/\psi\phi$
channel due to the near coincidence of the $D^{\pm}_{s}D^{\ast\pm}_{s}$ and $J/\psi\phi$ mass thresholds when the
nonrelativistic constituent quark model was employed~\cite{14}. According to the QCD sum rules,
Chen {\em et al.} pointed out that the $X(4500)$ and $X(4700)$ may be interpreted as the $D$-wave $cs\bar c\bar s$
tetraquark states with opposite color structures~\cite{15}. It should be emphasized that most of these explanations
do not agree with each other, and many of these investigations neglected the spin-orbital interaction when the orbital
excitation are taken into account.

To see whether these exotic resonances can be described by $cs\bar c\bar s$ tetraquark systems with $J^{PC}=0^{++},~1^{++}$,
we do a high precision four-body calculation based on the framework of the chiral quark model, which described the hadron
spectra and hadron-hadron interaction well~\cite{16,16a}. The high precision few-body method, Gaussian expansion method
(GEM)~\cite{20}, is employed for this purpose.
Two configurations, $(q\bar q)(q\bar q)$ (meson-meson) and $(qq)(\bar q\bar q)$ (diquark-antidiquark) are considered.
All the color constructions for each configuration are taken into account. For meson-meson
configuration, the color structures are $1\otimes1$ and $8\otimes8$, and for diquark-antidiquark configuration,
$\bar 3\otimes3$ and $6\otimes\bar 6$.
To explain the two higher exotic resonances $X(4500)$ and $X(4700)$, the orbital excitation with the inclusion of
spin-orbital interaction are invoked. To expose the structures of the states, the distances between two quarks (antiquarks)
for given states are calculated.

This paper is organized as follows. In Sec. \ref{WAVE FUNCTIONS}, the chiral quark model and wave functions of
$cs\bar c\bar s$ tetraquark systems are introduced. The numerical results with discussion are presented in
Sec. \ref{DISCUSSIONS}. Finally, we give a brief summary of our investigation and the future work to be done
in Sec. \ref{SUMMARY}.

\section{CHIRAL QUARK MODEL AND WAVE FUNCTIONS\label{WAVE FUNCTIONS}}
The chiral quark model has achieved great success when describing hadron spectra and hadron-hadron
interactions~\cite{16a}.
The specific introduction of the chiral quark model can be found in Ref. \cite{16}. The Hamiltonian of
$cs\bar c\bar s$ tetraquark systems, which is shown below, includes the mass, the kinetic energy and
different kinds of interactions. These interactions include the confinement $V^{C}$, one-gluon-exchange
$V^{G}$ and Goldstone bosons exchanges $V^{\chi}(\chi=\pi,\kappa,\eta)$, only $\eta$ exchange plays a role 
between $s$ and $\bar{s}$.
Scalar meson exchange $V^{\sigma}$ is not included in these interactions because it is expected to exist between
$u(\bar u)$ and $d(\bar d)$ only. Due to the existence of orbital excitation, the spin-orbit coupling terms are
also taken into consideration.
\begin{widetext}
\begin{eqnarray}
&&H=\sum_{i=1}^4(m_i+\frac{p_i^2}{2m_i})-T_{cm}
+\sum_{i=1<j}^4\left(V_{ij}^{C}+V_{ij}^{G}+\sum_{\chi=\pi,\kappa,\eta}V_{ij}^{\chi}\right)+V_{Q\bar Q}^{C,LS}+V_{Q\bar Q}^{G,LS}, \label{Eq.5}\\
&&V_{ij}^{C}= ( -a_c r_{ij}^2-\Delta ) \boldsymbol{\lambda}_i^c \cdot
 \boldsymbol{\lambda}_j^c \\
&&V_{ij}^{G}= \frac{\alpha_s}{4} \boldsymbol{\lambda}_i^c \cdot
\boldsymbol{\lambda}_{j}^c
\left[\frac{1}{r_{ij}}-\frac{2\pi}{3m_im_j}\boldsymbol{\sigma}_i\cdot
\boldsymbol{\sigma}_j
  \delta(\boldsymbol{r}_{ij})\right], \\
&&V_{ij}^{\pi} = {\frac{g_{ch}^{2}}{{4\pi
}}}{\frac{m_{\pi}^{2}}{{12m_{i}m_{j}}}}
{\frac{\Lambda _{\pi}^{2}~m_{\pi}}{{\Lambda _{\pi}^{2}-m_{\pi}^{2}}}}%
 \boldsymbol{\sigma}_{i}\cdot \boldsymbol{\sigma}_{j}\left[ Y(m_{\pi}\,r_{ij})-{\frac{\Lambda _{\pi}^{3}}{m_{\pi}^{3}}}%
Y(\Lambda _{\pi}r_{ij})\right]\sum_{a=1}^3\boldsymbol{\lambda}_{i}^{a}\cdot
\boldsymbol{\lambda}_{j}^{a}, \\
&&V_{ij}^{\kappa} = {\frac{g_{ch}^{2}}{{4\pi }}}{\frac{m_{\kappa}^{2}}{{\
12m_{i}m_{j}}}}{\frac{\Lambda _{\kappa}^{2}~m_{\kappa}}{{\Lambda _{\kappa}^{2}-m_{\kappa}^{2}}}}%
\boldsymbol{\sigma}_{i}\cdot \boldsymbol{\sigma}_{j}
\left[ Y(m_{\kappa}\,r_{ij})-{\frac{\Lambda _{\kappa}^{3}}{m_{\kappa}^{3}}}%
Y(\Lambda _{\kappa}r_{ij})\right]\sum_{a=4}^7\boldsymbol{\lambda}_{i}^{a}\cdot
\boldsymbol{\lambda}_{j}^{a}, \\
&&V_{ij}^{\eta} = {\frac{g_{ch}^{2}}{{4\pi
}}}{\frac{m_{\eta}^{2}}{{
12m_{i}m_{j}}}}{\frac{\Lambda _{\eta}^{2}}{{\Lambda _{\eta}^{2}-m_{\eta}^{2}}}}%
m_{\eta}  \boldsymbol{\sigma}_{i}\cdot \boldsymbol{\sigma}_{j}
\left[ Y(m_{\eta}\,r_{ij})-{\frac{\Lambda _{\eta}^{3}}{m_{\eta}^{3}}}%
Y(\Lambda _{\eta}r_{ij})\right]
\left[\cos\theta_P(\boldsymbol{\lambda}_{i}^{8}\cdot
\boldsymbol{\lambda}_{j}^{8})-\sin\theta_P(\boldsymbol{\lambda}_{i}^{0}\cdot \boldsymbol{\lambda}_{j}^{0})\right].
\end{eqnarray}
\end{widetext}
Where $T_{cm}$ is the kinetic energy of the center of mass motion; $Y(x)$ is the standard Yukawa functions;
the $\boldsymbol{\sigma}$ and $\boldsymbol{\lambda}$ represent Pauli and Gell-Mann matrices, respectively;
and the strong coupling constant of one-gluon exchange is $\alpha_s$, its running property is given as:
\begin{eqnarray}
\alpha_s(\mu_{ij})=\frac{\alpha_0}{\ln\left[(\mu_{ij}^2+\mu_0^2)/\Lambda_0^2\right]},
\end{eqnarray}
where $\mu_{ij}$ represents the reduced mass of two interacting particles.

For the diquark-antidiquark configuration, the sub-clusters $qq$ and $\bar q\bar q$ can be treated as
compound bosons $\bar Q$ and $Q$ with no internal orbital excitation. If the relative orbital angular excitations
between the two clusters is $\boldsymbol{L}$, the four-body spin-orbit interactions can be simply expressed
as follows~\cite{17,18}:
\begin{eqnarray}
&&V_{Q\bar Q}^{C,LS}=-a_c\boldsymbol{\lambda}_Q^c \cdot\boldsymbol{\lambda}_{\bar Q}^c\frac{1}{4M_Q M_{\bar Q}}\boldsymbol{L}\cdot\boldsymbol{S}\\
&&V_{Q\bar Q}^{G,LS}=-\frac{\alpha_s}{4} \boldsymbol{\lambda}_Q^c \cdot\boldsymbol{\lambda}_{\bar Q}^c\frac{1}{8M_Q M_{\bar Q}}\frac{3}{\boldsymbol{X}^3}\boldsymbol{L}\cdot\boldsymbol{S},
\end{eqnarray}
where the $M_{Q}(M_{\bar Q})$ is the total mass of sub-cluster; $\boldsymbol{X}$ is the distance between the two clusters,
and $\boldsymbol{S}$ is the total spin of the tetraquark state. This simplification can be generalized to
color-octet meson sub-clusters in meson-meson configuration.

In this paper, the model parameters of chiral quark model are directly taken from our previous work~\cite{19},
which is shown in Table~\ref{Table.1}. These parameters are obtained by fitting the meson spectrum. Some of the
calculated meson spectral have been given in Table~\ref{Table.2}.

Next, we will introduce the wave functions for $cs\bar c\bar s$ tetraquark.
A quark (antiquark) has four degrees of freedom, including orbit, flavor, spin, and color.
For each degree of freedom, first we construct the wave function for each sub-cluster, then coupling the wave functions
of two sub-clusters to get the wave functions for the final tetraquark systems.

For spatial part, the total spatial wave functions of tetraquark systems can be obtained by coupling
three relative orbital motion wave functions:
\begin{eqnarray}
\psi_{LM_{L}}=\left[\left[\psi_{l_1}(\mathbf{r}_{12})\psi_{l_2}(\mathbf{r}_{34})\right]_{l_{12}}
\psi_{L_r}(\mathbf{R})\right]_{LM_{L}}, \label{orbitalwf}
\end{eqnarray}
where $\psi_{l_1}(\mathbf{r}_{12})$ and $\psi_{l_1}(\mathbf{r}_{34})$ are the relative orbital motions
between two particles in each sub-cluster with angular momentum $l_1$ and $l_2$, respectively; and
$\psi_{L_r}(\mathbf{R})$ is the relative orbital wave function between two sub-clusters with angular momentum
$L_r$. In the present calculation, we do not consider the orbital excitation in each sub-cluster, so set $l_1=l_2=0$
and $L_r=L$, which is the total orbital angular momentum of tetraquark systems. In GEM,
three relative orbital motion wave functions are all expanded by Gaussian functions~\cite{20}:
\begin{eqnarray}
\psi_{lm}(\mathbf{r})=\sum^{n_{max}}_{n=1}c_{nl}\phi^{G}_{nlm}(\mathbf{r})
\end{eqnarray}
\begin{eqnarray}
\phi^{G}_{nlm}(\mathbf{r})=\emph{N}_{nl}r^{l}e^{-\nu_{n}r^{2}}\emph{Y}_{lm}(\hat{\mathbf{r}})
\end{eqnarray}
\begin{eqnarray}
\emph{N}_{nl}=\left(\frac{2^{l+2}(2\nu_{n})^{l+3/2}}{\sqrt\pi(2l+1)!!}\right)^{\frac{1}{2}},
\end{eqnarray}
where $N_{nl}$ are normalization constants, and the expansion coefficients $c_{nl}$ are obtained
by solving the Schr\"{o}dinger equation. The Gaussian size parameters are set according to the following
geometric progression:
\begin{eqnarray}
\nu_{n}=\frac{1}{r^{2}_{n}},r_{n}=r_{min}a^{n-1},a=\left(\frac{r_{max}}{r_{min}}\right)^{\frac{1}{n_{max}-1}},
\end{eqnarray}
where the $n_{max}$ is the number of Gaussian functions, and $a$ is the ratio coefficient. After parameter
optimization through continuous calculation, we found the calculation results begin to converge, when $n_{max}=7$.
\begin{table}
\caption{Parameters of chiral quark model.\label{Table.1}}
\begin{ruledtabular}
\begin{tabular}{llll}
Parameters&Values&Parameters&Values\\
\hline
$m_{u}$(MeV)&313&$\Lambda_{\pi}=\Lambda_{\sigma}(fm^{-1})$&4.2\\
$m_{d}$(MeV)&313&$\Lambda_{\eta}=\Lambda_{\kappa}(fm^{-1})$&5.2\\
$m_{s}$(MeV)&536&$g_{ch}^2/(4\pi)$&0.54\\
$m_{c}$(MeV)&1728&$\theta_p(^\circ)$&-15\\
$m_{b}$(MeV)&5112&$a_c$(MeV)&101\\
$m_{\pi}(fm^{-1})$&0.70&$\Delta$(MeV)&-78.3\\
$m_{\sigma}(fm^{-1})$&3.42&$\alpha_0$&3.67\\
$m_{\eta}(fm^{-1})$&2.77&$\Lambda_0(fm^{-1})$&0.033\\
$m_{\kappa}(fm^{-1})$&2.51&$\mu_0$(MeV)&36.976\\
\end{tabular}
\end{ruledtabular}
\end{table}
\begin{table}
\caption{Meson spectra (unit: MeV).\label{Table.2}}
\begin{ruledtabular}
\begin{tabular}{ccrccr}
mesons & $E$ & PDG~\cite{PDG} & mesons & $E$ & PDG~\cite{PDG} \\
\hline
$\pi$&140.1&139.6&$D_{s}$&1953.4&1968.3\\
$\rho$&774.4&775.3&$D_{s}^{\star}$&2080.2&2112.2\\
$\omega$&708.2&782.7&$\phi$&1015.8&1019.5\\
$K$&496.4&493.7&$\eta^{\prime}$&824.0&957.8\\
$K^{\star}$&918.4&891.8&$\eta_{c}$&2986.3&2983.9\\
$D$&1875.4&1869.7&$J/\psi$&3096.4&3096.9\\
$D^{\star}$&1986.3&2010.3&&&\\
\end{tabular}
\end{ruledtabular}
\end{table}

For spin part, since the spin of each quark (antiquark) is 1/2, the spin of each sub-cluster can only be 0 or 1.
After coupling the spins of the two sub-clusters, the total spin of tetraquark may be 0, 1, and 2.
The total spin $S$ of tetraquark system is obtained from the coupling: $S_1\otimes S_2\rightarrow S$,
where $S_1$ and $S_2$ represent the spins of two sub-clusters. We use $\chi_i(i=1\sim6)$ to denote the total
spin wave functions of tetraquark systems. All of the six possible spin channels are given as:
\begin{eqnarray}
\chi_{1}:0\otimes0\rightarrow0 &
\chi_{2}:1\otimes1\rightarrow0 & \nonumber \\
\chi_{3}:0\otimes1\rightarrow1 &
\chi_{4}:1\otimes0\rightarrow1 &
\chi_{5}:1\otimes1\rightarrow1 \\
\chi_{6}:1\otimes1\rightarrow2 & & \nonumber
\end{eqnarray}

For flavor part, the isospins of all quarks (antiquarks) are all zero, so there is no need to consider
the coupling of the isospin. We use $\varphi_j(j=1\sim3)$ to represent the total flavor wave functions of
tetraquark systems and they can be written as:
\begin{eqnarray}
\varphi_{1} & = & (c\bar c)(s\bar s), \nonumber \\
\varphi_{2} & = & (c\bar s)(s\bar c), \\
\varphi_{3} & = & (cs)(\bar c\bar s), \nonumber
\end{eqnarray}
where $\varphi_{1}$ and $\varphi_{2}$ are for meson-meson configuration and $\varphi_{3}$ is for
diquark-antidiquark configuration.

For color part, there are four different color wave functions of tetraquark systems, which are denoted by
$\omega_k~(k=1\sim4)$. $\omega_{1}$ and $\omega_{2}$ stand for the color singlet-singlet $1\otimes 1$ and
octet-octet $8\otimes 8$ in meson-meson configuration, respectively. The remaining two color wave functions
$\omega_{3}$ and $\omega_{4}$ stand for the color antitriplet-triplet $ \bar 3\otimes 3$ and sextet-antisextet
$6\otimes \bar{6}$ in diquark-antidiquark configuration, respectively. The specific construction process and
forms of these color wave functions can be found in Ref. \cite{19}.

Finally, the total wave functions for the tetraquark systems can be given as:
\begin{eqnarray}
&&\Psi_{LJM_{J}}^{ijk}={\cal A}[\psi_{L}\chi_{i}]^{J}_{M_{J}}\varphi_{j}\omega_{k}, \label{totalwf}\\
&&(i=1\sim6,~j=1\sim3,~k=1\sim4), \nonumber
\end{eqnarray}
where $J$ is the total angular momentum and $M_{J}$ is the 3rd component of the total angular momentum.
Due to the two quarks (antiquarks) in $cs\bar c\bar s$ tetraquark systems are all
non-identical particles, the the antisymmetrization operator ${\cal A}=1$.

The eigenenergies of the tetraquark systems can be obtained by solving the following Schr\"odinger equation:
\begin{eqnarray}
H\Psi_{LJM_{J}}^{ijk}=E_{LJM_{J}}^{ijk}~\Psi_{LJM_{J}}^{ijk},
\end{eqnarray}
where the Hamiltonian $H$ and wave functions $\Psi_{LJM_{J}}^{ijk}$ have been given in Eq.(\ref{Eq.5}) and
Eq.(\ref{totalwf}), respectively.

\section{NUMERICAL RESULTS AND DISCUSSIONS\label{DISCUSSIONS}}
\begin{table*}
\caption{Energy of $S$-wave $cs\bar c\bar s$ tetraquark states with $J^{PC}=1^{++}$ (unit: MeV).\label{Table3}}
\begin{ruledtabular}
\begin{tabular}{cccccc}
Channel&$E_{1S}$&$E_{c1}$&$E^{\prime}_{c1}$&$E^{theo}_{th}$&$E^{exp}_{th}$\\
\hline
$\psi_S\chi_5\varphi_1\omega_1$&4112.7&4112.7&4116.9&4112.2&4116.4$(J/\psi\phi)$\\
$\psi_S\chi_5\varphi_1\omega_2$&4305.2&4305.2&4309.4&&            \\
$\psi_S\chi_p\varphi_2\omega_1$&4033.8&4033.7&4080.7&4033.5&4080.5$(D_{s}D^{\ast}_{s})$\\
$\psi_S\chi_p\varphi_2\omega_2$&4370.6&4370.8&4417.8&&            \\
$\psi_S\chi_p\varphi_3\omega_3$&4343.4&4332.7&-&&\\
$\psi_S\chi_p\varphi_3\omega_4$&4361.5&4374.3&-&&            \\
\end{tabular}
\end{ruledtabular}
\end{table*}
\begin{table*}
\caption{Energy of $S$-wave $cs\bar c\bar s$ tetraquark states with $J^{PC}=0^{++}$ (unit: MeV).\label{Table4}}
\begin{ruledtabular}
\begin{tabular}{ccccccccc}
Channel&$E_{1S}$&$E_{c1}$&$E_{c1}^{\prime}$&$E^{theo}_{th}$&$E^{exp}_{th}$&$E_{2S}$&$E_{c2}$&$E_{c2}^{\prime}$\\
\hline
$\psi_S\chi_1\varphi_1\omega_1$&3810.7&3810.6& 3942.0&3810.3&3941.7($\eta_{c}\eta^{\prime}$)&&&     \\
$\psi_S\chi_1\varphi_1\omega_2$&4359.1&4360.2&4491.6&&&4712.7&&4844.1   \\
$\psi_S\chi_2\varphi_1\omega_1$&4114.0&4114.0&4118.2&4112.2&$4116.4(J/\psi\phi)$&&&       \\
$\psi_S\chi_2\varphi_1\omega_2$&4273.4&4273.4&4277.6&&&4601.2&&4605.4       \\
$\psi_S\chi_1\varphi_2\omega_1$&3908.0&3908.0&3837.8&3906.8&$3936.6(D_{s}D_{s})$&&&       \\
$\psi_S\chi_1\varphi_2\omega_2$&4376.1&4376.3&4406.1&&&4910.6&&4940.4       \\
$\psi_S\chi_2\varphi_2\omega_1$&4161.6&4161.6&4225.6&4160.4&$4224.4(D^{\ast}_{s}D^{\ast}_{s})$&&&       \\
$\psi_S\chi_2\varphi_2\omega_2$&4304.8&4304.4&4368.4&&&4655.4&&4719.4      \\
$\psi_S\chi_1\varphi_3\omega_3$&4323.7&4319.1&-&&&4643.5&4641.7&-       \\
$\psi_S\chi_1\varphi_3\omega_4$&4384.9&4390.0&-&&&4976.5&4980.4&-      \\
$\psi_S\chi_2\varphi_3\omega_3$&4370.0&4376.1&-&&&4924.3&4926.8&-       \\
$\psi_S\chi_2\varphi_3\omega_4$&4304.2&4292.1&-&&&4607.4&4601.4&-       \\
\end{tabular}
\end{ruledtabular}
\end{table*}
The $J^{PC}$ of these exotic resonances observed recently in $B^+\rightarrow J/\psi\phi K^+$ decay
are fixed to $1^{++}$ and $0^{++}$. In our current calculation, only the states with these two sets of quantum
numbers are considered. Due to the parities of these exotic resonances is positive, the total orbital angular momentum
$L$ must be even, so does $L_r$, according to Eq.(\ref{orbitalwf}).
Meanwhile, these exotic resonances all have the definite positive $C$ parity, we need to combine the spin and flavor
wave functions to construct the eigenstates of the charge conjugate operator.
We use $[(q\bar q)^{S_1}(q\bar q)^{S_2}]^S$ and $[(qq)^{S_1}(\bar q\bar q)^{S_2}]^S$ to represent the combination of
spin and flavor wave functions. According to the Ref. \cite{21}, all of these eigenstates with positive $C$ parity
are given as:
\begin{eqnarray}
&&\chi_{5}\varphi_{1}=[(c\bar c)^1(s\bar s)^1]^1,\\
&&\chi_{p}\varphi_{2}=\frac{1}{\sqrt{2}}([(c\bar s)^0(s\bar c)^1]^1+
[(c\bar s)^1(s\bar c)^0]^1),\label{Eq.31}\\
&&\chi_{p}\varphi_{3}=\frac{1}{\sqrt{2}}([(cs)^0(\bar c\bar s)^1]^1+
[(cs)^1(\bar c\bar s)^0]^1),\label{Eq.32}\\
&&\chi_{1}\varphi_{1}=[(c\bar c)^0(s\bar s)^0]^0,~~~\chi_{2}\varphi_{1}=[(c\bar c)^1(s\bar s)^1]^0,\\
&&\chi_{1}\varphi_{2}=[(c\bar s)^0(s\bar c)^0]^0,~~~\chi_{2}\varphi_{2}=[(c\bar s)^1(s\bar c)^1]^0,\\
&&\chi_{1}\varphi_{3}=[(cs)^0(\bar c\bar s)^0]^0,~~~\chi_{2}\varphi_{3}=[(cs)^1(\bar c\bar s)^1]^0,\\
&&\chi_{6}\varphi_{1}=[(c\bar c)^1(s\bar s)^1]^2\\
&&\chi_{6}\varphi_{2}=[(c\bar s)^1(s\bar c)^1]^2\\
&&\chi_{6}\varphi_{3}=[(cs)^1(\bar c\bar s)^1]^2
\end{eqnarray}
where the $\chi_{p}$ in Eq.(\ref{Eq.31}) and Eq.(\ref{Eq.32}) is $\frac{1}{\sqrt{2}}(\chi_{3}+\chi_{4})$.

Generally, we only consider the $S$-wave states for the low-lying state, i.e., $L_r=0$ for states with
$J^{PC}=1^{++}$. However, for the high-lying states, $X(4500)$ and $X(4700)$, we will take into account the
$D$-wave states, i.e., $L_r=2$ for some states with $J^{PC}=0^{++}$.

First, we calculated the energies of the $S$-wave $cs\bar c\bar s$ states with $J^{PC}=1^{++}$ and
$0^{++}$. The results are given in Table~\ref{Table3} and Table~\ref{Table4}, respectively. In the tables,
$E_{1S}$ ($E_{2S}$) denotes the energy of the first (second) $S$-wave state
with single channel calculation, and $E_{c1}~(E_{c2})$ represents the energy through coupling of two different color
structures. $E^{theo}_{th}~(E^{exp}_{th})$ represents the theoretical (experimental) two-body threshold.
Because the theoretical calculation cannot reproduced the experimental data exactly for meson spectrum, so
we make a correction, $E^{\prime}= E-E^{theo}_{th}+E^{exp}_{th}$ for the meson-meson configuration to
minimize the theoretical uncertainty.
$E_{c1}^{\prime}~(E_{c2}^{\prime})$ represents the corrected energy of $E_{c1}~(E_{c2})$.
For diquark-antidiquark configuration, the correction is not applied because no asymptotic physical state in this case.
\begin{figure*}
\centering
\includegraphics[width=8cm]{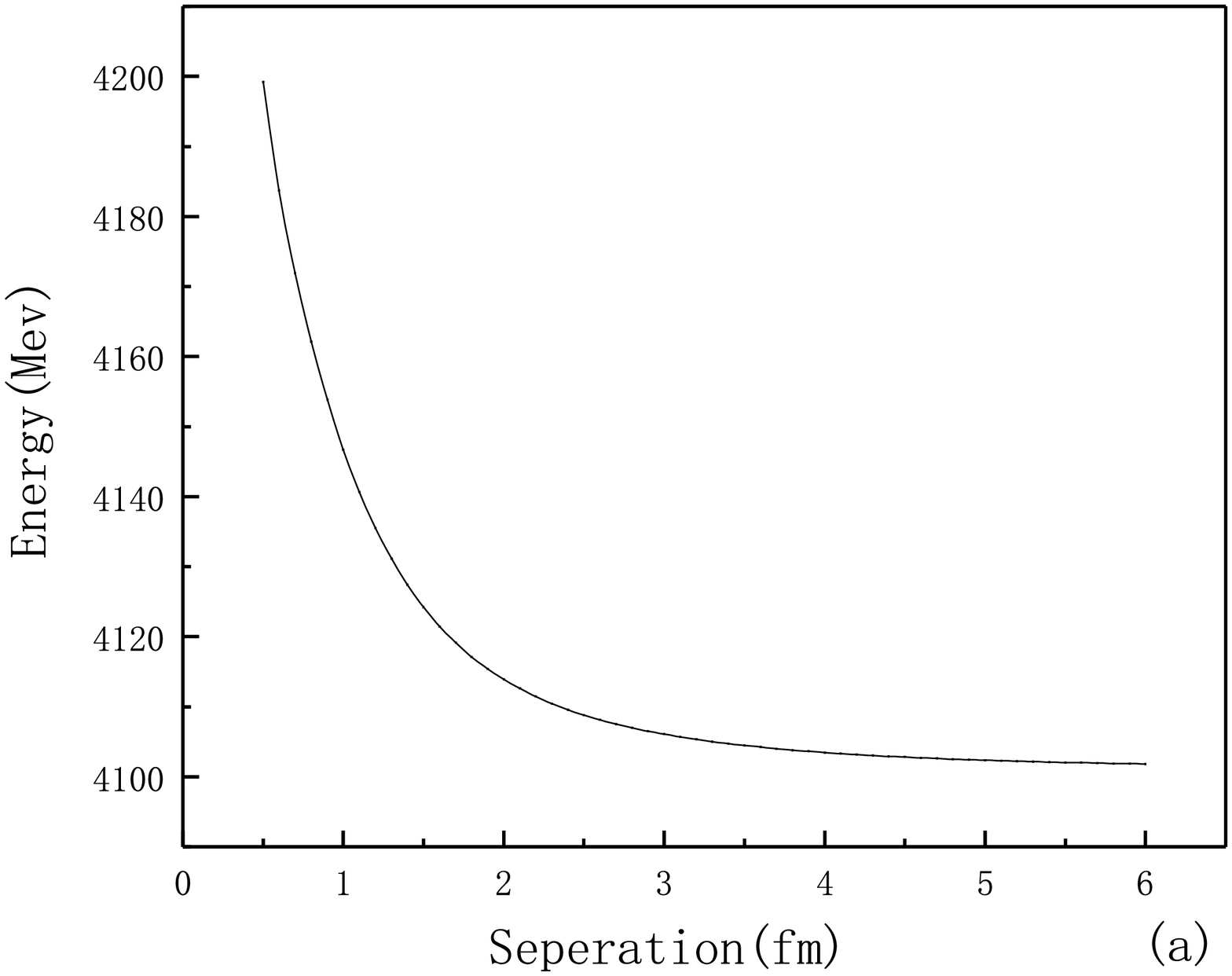}
\includegraphics[width=8cm]{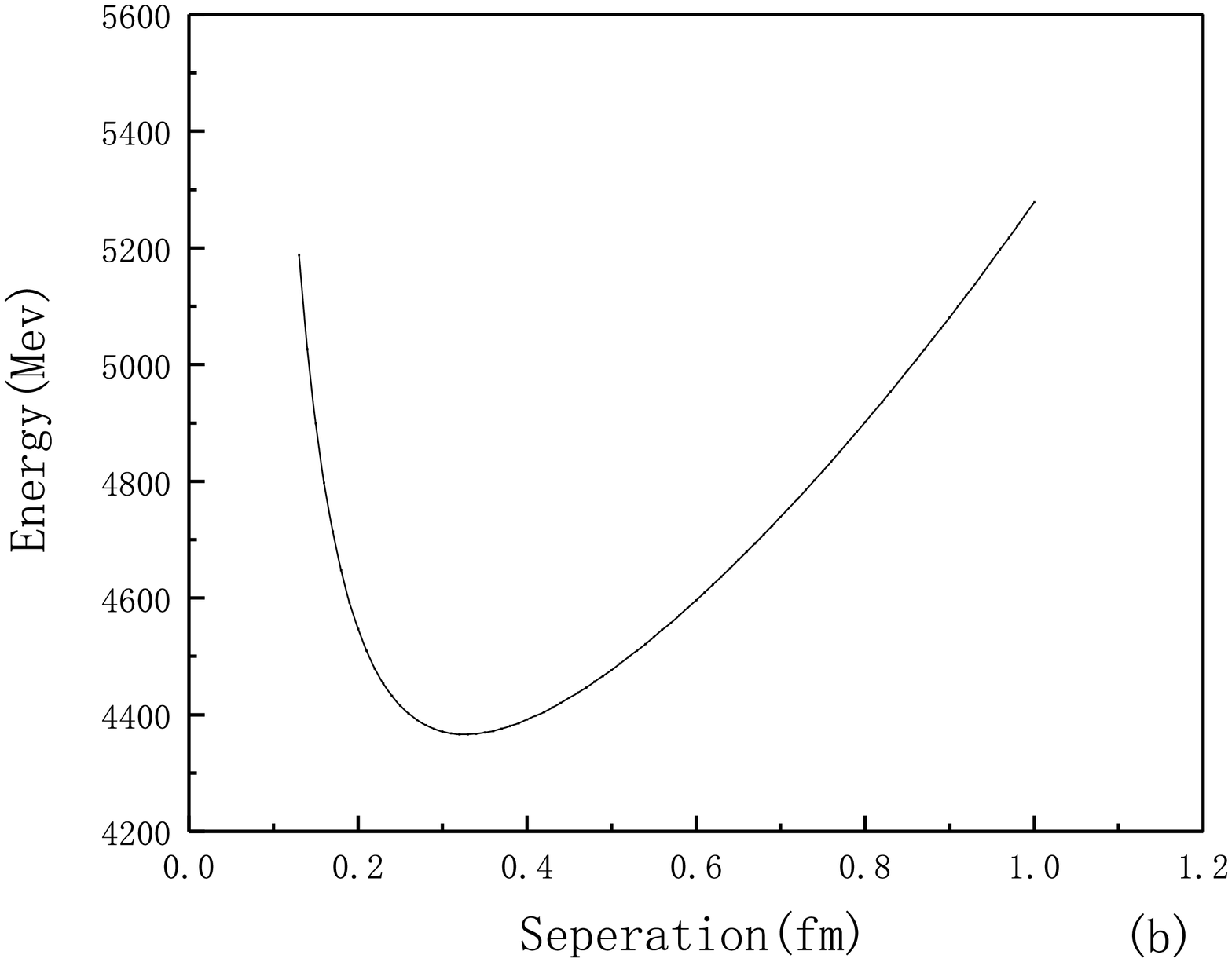}
\includegraphics[width=8cm]{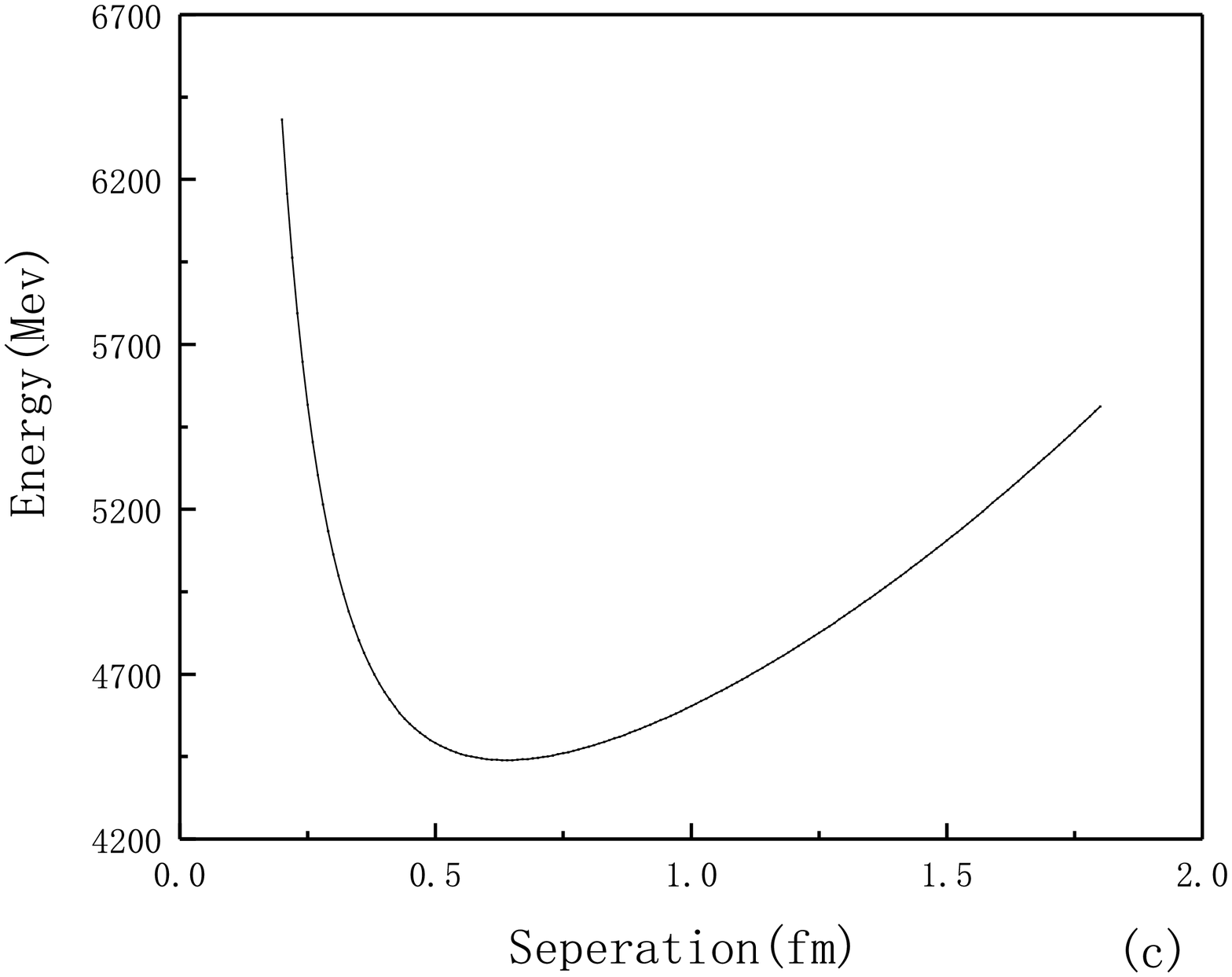}
\includegraphics[width=8cm]{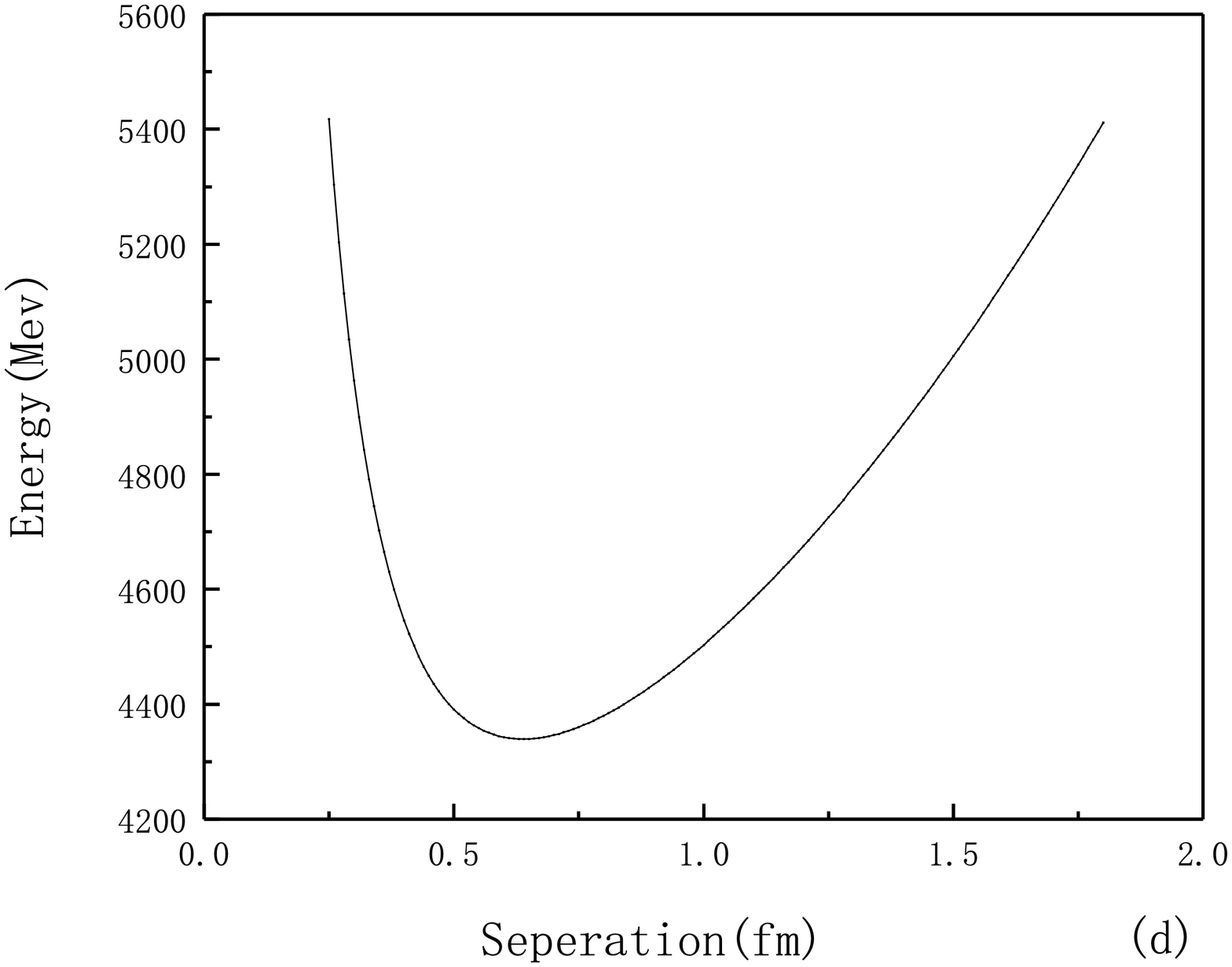}
\caption{Energy of $cs\bar c\bar s$ tetraquark states with different color structures as a function of the distance between two sub-clusters. a: $1\otimes1$, b: $8\otimes8$, c: $\bar 3\otimes3$ and d: $6\otimes\bar 6$}
\end{figure*}

From Table~\ref{Table3} and Table~\ref{Table4}, we can see that the energies of color singlet-singlet
$(1\otimes1)$ structure are all a bit high than the corresponding theoretical thresholds. The adiabatic energy
of the system in this case is shown in Fig. 1 (the adiabatic energy is obtained by setting the number of Gaussians
for the relative motion between two sub-clusters to 1). When we gradually increase the distance between the two
sub-clusters, the energy slowly tend to theoretical threshold. This phenomenon suggests
that the two mesons tend to stay away and no bound states can be formed when the color structure is $1\otimes1$.
The reason for this phenomenon is that the Goldstone bosons exchange between the two sub-clusters is too weak
to bind the two mesons together. When the color structure is octet-octet $(8\otimes8)$, the energies
are generally higher than the corresponding color singlet-singlet $(1\otimes1)$ structure.
According to Fig.1(b), the two colorful sub-clusters cannot fall apart or get too close. That's because the existence 
of confinement $V_{C}$ hinders the too far separation of the two colorful clusters. This phenomenon suggests that the resonances
with compact tetraquark structure may exist in our present calculation when the color configuration of the meson-meson
configuration is octet-octet $(8\otimes8)$. Due to the small overlap between the two color structures, $1\otimes 1$ and
$8\otimes8$, the coupling between two color structures is small, which makes the resonance possible.
When the configuration is diquark-antidiquark, the two color structures antitriplet-triplet $(\bar 3\otimes3)$ and
sextet-antisextet$ (6\otimes\bar 6)$ have the similar energies and the coupling between them is rather strong.  
According to Fig.1(c) and Fig.1(d),
whether $\bar 3\otimes3$ or $6\otimes\bar 6$, the two sub-clusters are all colorful and they cannot stay too
far away from each other, and short-range repulsion prevent the two sub-clusters from getting too close. 
So the resonances with compact tetraquark structure may also exist in diquark-antidiquark configuration.
\begin{table*}
\caption{Energy of $D$-wave $cs\bar c\bar s$ tetraquark states with $J^{PC}=0^{++}$.(unit:Mev)\label{Table5}}
\begin{ruledtabular}
\begin{tabular}{cccccc}
Channel&$E_{1D}$&$E_{c1}$&$E_{c1}^{\prime}$&$E^{theo}_{th}$&$E^{exp}_{th}$ \\
\hline
$[\psi_D\chi_6]^0\varphi_1\omega_1$&4116.6&4116.4&4120.6&4112.2&4116.4$(J/\psi\phi)$\\
$[\psi_D\chi_6]^0\varphi_1\omega_2$&5173.8&5174.0&5178.2&&            \\
$[\psi_D\chi_6]^0\varphi_2\omega_1$&4163.5&4163.2&4227.2&4160.4&$4224.4(D^{\ast}_{s}D^{\ast}_{s})$\\
$[\psi_D\chi_6]^0\varphi_2\omega_2$&5141.0&5141.0&5205.0&&            \\
$[\psi_D\chi_6]^0\varphi_3\omega_3$&4853.8&4847.6&-&\\
$[\psi_D\chi_6]^0\varphi_3\omega_4$&5173.8&5177.6&-&            \\
\end{tabular}
\end{ruledtabular}
\end{table*}

\begin{table}
\caption{Distance between $q(\bar q)$ and $q(\bar q)$ for $(c\bar s)(s\bar c)$ ground states with $J^{PC}=0^{++}$.\label{Table6}}
\begin{ruledtabular}
\begin{tabular}{ccccccc}
Channel&$c\bar s$&$s\bar c$&$cs$&$\bar s\bar c$&$c\bar c$&$\bar ss$ \\
\hline
$\psi_S\chi_1\varphi_2\omega_1$&0.50&0.50&5.86&5.86&5.84&5.87            \\
$\psi_S\chi_1\varphi_2\omega_2$&0.70&0.70&0.67&0.67&0.44&0.84            \\
Coupling&0.50&0.50&5.86&5.86&5.84&5.87\\
&0.71&0.71&0.68&0.68&0.43&0.85\\
$\psi_S\chi_2\varphi_1\omega_1$&0.60&0.60&5.86&5.86&5.85&5.88            \\
$\psi_S\chi_2\varphi_1\omega_2$&0.67&0.67&0.64&0.64&0.42&0.80            \\
Coupling&0.60&0.60&5.86&5.86&5.85&5.88\\
&0.69&0.69&0.68&0.68&0.46&0.85\\
\end{tabular}
\end{ruledtabular}
\end{table}
\begin{table}
\caption{Distance between $q(\bar q)$ and $q(\bar q)$ for $(cs)(\bar c\bar s)$ ground states with $J^{PC}=0^{++}$.\label{Table7}}
\begin{ruledtabular}
\begin{tabular}{ccccccc}
Channel&$cs$&$\bar c\bar s$&$c\bar c$&$s\bar s$&$c\bar s$&$s\bar c$ \\
\hline
$\psi_S\chi_1\varphi_3\omega_3$&0.61&0.61&0.51&0.80&0.67&0.67            \\
$\psi_S\chi_1\varphi_3\omega_4$&0.71&0.71&0.44&0.85&0.68&0.68            \\
Coupling&0.59&0.59&0.32&0.69&0.54&0.54\\
&0.75&0.75&0.57&0.96&0.79&0.79\\
$\psi_S\chi_2\varphi_3\omega_3$&0.63&0.63&0.50&0.82&0.68&0.68            \\
$\psi_S\chi_2\varphi_3\omega_4$&0.67&0.67&0.41&0.80&0.63&0.63            \\
Coupling&0.42&0.42&0.37&0.56&0.48&0.48\\
&0.83&0.83&0.52&0.99&0.79&0.79\\
\end{tabular}
\end{ruledtabular}
\end{table}
For the states with $J^{PC}=1^{++}$ (Table~\ref{Table3}), the energies of $1S$ ground state are all between 4300 MeV
and 4420 MeV except the state with color structure $1\otimes1$. The corrected energy $E_{c1}^{\prime}$ of hidden
color($8\otimes 8$) channel is around 4309.4 MeV and the lowest energy of diquark-antidiquark configuration is about 4332 MeV.
Both energies are not far from the mass of $X(4274)$. The mix of two configurations may reduced the energy
a little, the lowest energy of the state with $J^{PC}=1^{++}$ is expected to approach the experimental value of the
$X(4274)$. So the $X(4274)$ can be a candidate of compact tetraquark state in our present calculation.
As for $X(4140)$, the energy is too low and there is no matching state for it in our calculations.
For the states with $J^{PC}=0^{++}$ (Table~\ref{Table4}), the energies of $1S$ ground state are all between
4277 MeV and 4492 MeV except the state with color structure $1\otimes1$. It is worth mentioning that the energy of
the $X(4350)$, which reported by Belle Collaboration in $\gamma\gamma\rightarrow J/\psi\phi$ process, is close to
the energy of hidden color($8\otimes 8$) state of $(c\bar{s})(s\bar{c})$, 4368.4 MeV. If the $J^{PC}$ of $X(4350)$ can be determined
to be $0^{++}$, we think it can be a candidate of compact tetraquark state.
As for the $2S$ radial excitation, the energy of hidden color($8\otimes 8$) state of $(c\bar{s})(s\bar{c})$ is 4719.4 MeV, which is
very close to the mass of the $X(4700)$ found by LHCb Collaboration.
In addition to the hidden-color state with color structure $8\otimes 8$, the $2S$ states with diquark-antidiquark
configuration have energies around 4650 MeV, which are not much different from the $X(4700)$, too.
So the $X(4700)$ can be explained to be a $2S$ radial excited state with compact tetraquark structure in our calculation.
The other exotic resonance recently found in experiment by LHCb Collaboration is $X(4500)$, but there is no matching state
for it in our calculations. Whether $X(4500)$ can be explained by tetraquark structure requires further study.

For the $D$-wave states with $J^{PC}=0^{++}$ (see Table~\ref{Table5}), the energies of meson-meson configuration
with color structure $1\otimes1$ are still very close to the corresponding theoretical threshold. That's because
the too long distance between the two clusters make the effect of the $D$-wave excitation negligible.
When the color structure is $8\otimes8$ or the configuration is diquark-antidiquark, the energies of $D$-wave
excitation are all too high to be explained as the $X(4700)$ and $X(4500)$. So it's not appropriate to use angular
excitation to explain these exotic resonances in our present calculation.

In order to analyse the spacial structure of the $cs\bar c\bar s$ tetraquark, we calculated the distances between two 
$q(\bar q)$ for $(c\bar s)(s\bar c)$ and $(cs)(\bar c\bar s)$ ground states with $J^{PC}=0^{++}$ (see Table \ref{Table6} 
and Table \ref{Table7}). According to Table \ref{Table6}, when the color structure is $1\otimes1$, the distance 
between the two $q(\bar q)$ which are in two different sub-clusters is big. This phenomenon shows that two sub-clusters 
tend to stay away from each other. When the color structure is $8\otimes8$, the distances between any two $q(\bar q)$ 
are all very close and now the structure is a compact tetraquark structure.
After coupling the two color structures for meson-meson configuration, the change is small,
because of the weak coupling between these two color structures.
For the diquark-antiquark configuration in Table \ref{Table7}, whatever the color structure, the distance between any two 
$q(\bar q)$ is small and now the structure is also very compact. After coupling the two color structures for 
diquark-antiquark configuration, the change is obvious. Strong coupling between these two color structures is 
the cause of this phenomenon. In addition, the relative kinetic energy between two light $q(\bar q)$ is greater than 
that of two heavy $q(\bar q)$. That's why the distance between two light $q(\bar q)$ is bigger than that of two heavy 
$q(\bar q)$ in our calculations.

\section{SUMMARY AND OUTLOOK\label{SUMMARY}}
In this work, we investigated the $cs\bar c\bar s$ tetraquark states in the framework of chiral quark model
and try to explain those exotic resonances recently observed in the invariant mass distribution of $J/\psi\phi$.
Two configurations, $(q\bar q)(q\bar q)$ and $(qq)(\bar q\bar q)$, with all possible spin and color
structures are taken into consideration. We found that the $(q\bar q)(q\bar q)$ configuration can not form the
bound states when color structure is $1\otimes1$ because the Goldstone bosons exchange is too weak to bound the
two mesons. If the color structure of $(q\bar q)(q\bar q)$ is $8\otimes8$ or the configuration is $(qq)(\bar q\bar q)$,
the resonances can be formed.
In our calculation, the $X(4274)$ and $X(4350)$ in experiment can be regarded as the ground state of $cs\bar c\bar s$
compact tetraquark states. The $X(4700)$ can be explained to be the $2S$ radial excitation but the $X(4500)$ has no
match in present calculation.
Due to the too low energy of $X(4140)$, it is impossible to use tetraquark state to explain this exotic resonances
in our work. The current work doesn't consider the coupling between the configurations $(q\bar q)(q\bar q)$ and
$(qq)(\bar q\bar q)$ and the coupling of $S$-wave and $D$-wave is also ignored. These legacy situations will be
considered in the future work.

\end{document}